# Direct Evidence of a Major Merger in the Perseus Cluster

Kim HyeongHan[1], M. James Jee[1,2*], Wonki Lee[1], John ZuHone[3], Irina Zhuravleva[4], Wooseok Kang[5,6], and Ho Seong Hwang[5,7,8]

[1]Department of Astronomy, Yonsei University, 50 Yonsei-ro, Seoul 03722, Korea.

[2]Department of Physics and Astronomy, University of California, Davis, One Shields Avenue, Davis, CA 95616, USA.

[3]Center for Astrophysics | Harvard & Smithsonian, 60 Garden Street, Cambridge, MA 02138, USA

[4]Department of Astronomy and Astrophysics, University of Chicago, Chicago, 5640 S Ellis Avenue, Chicago IL 60637, USA

[5]Department of Physics and Astronomy, Seoul National University, 1 Gwanak-ro, Gwanak-gu, Seoul 08826, Republic of Korea

[6]Department of Physics, Korea Advanced Institute of Science and Technology (KAIST), 291, Daehak-ro, Yuseong-gu, Daejeon 34141, Republic of Korea

[7]SNU Astronomy Research Center, Seoul National University, 1 Gwanak-ro, Gwanak-gu, Seoul 08826, Republic of Korea

[8]Australian Astronomical Optics – Macquarie University, 105 Delhi Road, North Ryde, NSW 2113, Australia

*e-mail: ahrtears54@yonsei.ac.kr, mkjee@yonsei.ac.kr

## Abstract

**Although the Perseus cluster has often been regarded as an archetypical relaxed galaxy cluster, several lines of evidence including ancient, large-scale cold fronts, asymmetric plasma morphology, filamentary galaxy distribution, etc., provide a conflicting view of its dynamical state, suggesting that the cluster might have experienced a major merger. However, the absence of a clear merging companion identified to date hampers our understanding of the evolutionary track of the Perseus cluster consistent with these observational features. In this paper, through careful weak lensing analysis, we successfully identified the missing subcluster halo ($M_{200} = 1.70^{+0.73}_{-0.59} \times 10^{14}\ M_\odot$) at the > 5σ level centered on NGC1264, which is located ~430 kpc west of the Perseus main cluster core. Moreover, a significant (>3σ) mass bridge, which is also traced by the cluster member galaxies, is detected**

**between the Perseus main and sub clusters, which serves as direct evidence of gravitational interaction. With idealized numerical simulations, we demonstrate that a ~3:1 off-axis major merger can create the cold front observed ~700 kpc east of the main cluster core and generate the observed mass bridge through multiple core crossings. This discovery resolves the long-standing puzzle of Perseus' dynamical state.**

**Introduction**

The Perseus cluster (A426; $z = 0.0179$) is the X-ray brightest galaxy cluster[1], exhibiting a large cooling flow centered on the brightest cluster galaxy (BCG) NGC1275[2], a radio mini-halo[3,4], and exemplary sloshing[5-8] features. These phenomena characterize the cluster as an archetypical relaxed cluster. However, the cluster also exhibits characteristics of a dynamically young cluster: 1) its member galaxies are distributed in the east-west direction[9,10], 2) the plasma shows an asymmetry toward the same direction[5], and 3) its distinct X-ray cold fronts extend to very large radii[7,8]. These pieces of evidence observed in its baryonic components indicates that perhaps the Perseus cluster has undergone a major merger. However, the absence of a clear merging companion identified to date presents a puzzle in our understanding of its evolutionary track.

In this study, we report a discovery of the missing merging companion of the Perseus cluster based on weak-lensing (WL) analysis. In addition, we identify a significant mass bridge (>3σ) connecting the main and sub clusters. Our idealized numerical simulations whose initial conditions are motivated by our WL results predict that a ~3:1 off-axis major merger can create the observed cold front ~700 kpc east of the main cluster core. We also find that a significant mass bridge can form from the merger and enhance its density through multiple core crossings.

We present the first WL analysis of the Perseus cluster with archival Subaru/Hyper Suprime-Cam[11-14] and Sloan Digital Sky Survey[15] (SDSS) imaging data, representing the lowest (z = 0.0179) redshift galaxy cluster ever measured with WL. Although low-redshift (z ≪ 0.1) galaxy clusters inherently suffer from low lensing efficiency, the net signal-to-noise ratio of the lensing signal per physical area at the cluster redshift is high[16]. We achieve a source density of ~22.7 per sq. arcmin. This corresponds to ~47,500 per sq. Mpc at the redshift of the Perseus cluster, which is higher by a factor of ~280 than the case if the target lens were at $z = 0.5$ (see Methods for details). Additionally, we benefit from the high purity (>99%) of a background galaxy population, as most faint galaxies in the observed field are behind the Perseus cluster. Moreover, the noise from the large-scale structure along the line of sight per physical area is reduced due to a larger angular extent of the cluster. Finally, the proximity of the Perseus cluster renders the lensing signal insensitive to source redshifts, thereby eliminating the need for precise source redshift estimation. These advantages enable us to detect low-contrast mass structures, as long as the observational WL systematic errors are under control.

**Results**

**Two-Dimensional Mass Distribution**

Figure 1 shows the reconstructed mass (white) distribution around the cluster center (R < 40′) based on the HSC shear catalog. The strongest convergence peak (> 9σ) is centered at the BCG, NGC 1275 (z = 0.01756), bridged to the other peak (> 5σ) associated with the spiral galaxy NGC 1264 (z = 0.01109) whose the line-of-sight velocity difference is ∼1,800 km s$^{-1}$. The galaxy density peak is also in good agreement with the main halo peak, while the overall galaxy distribution stretches along the east-west direction[9,10], aligning with the significant (>3σ) mass bridge indicating gravitational interaction in the past. We found that the current public XMM-Newton and Chandra data suggest no significant excess in X-ray emission at the second mass peak. With the spectroscopic catalog[17] of the Perseus field, we find that the line-of-sight velocities of the cluster galaxies of the substructures are consistent within the uncertainties, supporting a scenario wherein the merger might have occurred in the plane of the sky (Extended Data Figure 1). Moreover, we confirmed that there is no background galaxy overdensity associated with the second mass peak.

**Parametric Mass Estimation**

Figure 2 presents a reduced tangential shear profile out to the cluster virial radius measured from a joint analysis of the HSC and SDSS data. The profile is centered at the BCG as it spatially coincides well with the mass center. The data points within the central $5'$ (~110 kpc) radius are discarded to minimize effects of the centroid bias, contamination by member galaxies, and nonlinear shear responsivity. To estimate the cluster mass, we employ two kinds of halo density profiles: a Navarro-Frenk-White[18,19] (NFW) profile and a singular isothermal sphere (SIS) profile. In the case of the mass bridge, we assume a cylindrical geometry $\kappa(h) = \kappa_0/[1 + (h/h_c)^2]$[16], where $\kappa_0$ and $h_c$ are the central density and the characteristic width, respectively. We summarize our best-fit results of the cluster mass in Extended Data Table 1.

We estimate the virial mass of the cluster to be $M_{200} = 6.82 \pm 1.76 \times 10^{14}~M_\odot$ assuming a single NFW halo profile with a concentration-mass (*c-M*) relation[20]. This result shows excellent agreement with the mass obtained in the X-ray study[21] ($M_{200,X-ray} = 6.65 \times 10^{14}~M_\odot$). The SIS fit predicts a velocity dispersion of $\sigma_{WL} = 808 \pm 53~km~s^{-1}$, which is slightly lower than the direct (spectroscopic) measurement of $\sigma_v \sim 1,000~km~s^{-1}$ [10,17,22,23]. We further explore the *c-M* parameter space by conducting Markov Chain Monte Carlo (MCMC) analysis using a single NFW profile[24,25]. We set a broad range of flat prior for the concentration (mass) parameter ranging from $c$=0 ($M_{200} = 1 \times 10^{12}\ M_\odot$) to 30 ($1 \times 10^{16}\ M_\odot$). The marginalized posterior distributions yield a mass of $M_{200} = 2.89^{+1.95}_{-0.98} \times 10^{14}\ M_\odot$ and a concentration parameter of $c = 10.61^{+7.84}_{-4.62}$.

The substructures in the reconstructed mass map (Figure 1) motivate modeling the Perseus cluster with two subhalos. We estimate the masses of these substructures by simultaneously fitting two NFW halo density profiles[24,25,27]. We place each halo at the location of the nearest brightest galaxy, which is in good spatial agreement with the mass peak (Figure 1). Choosing different halo centers, such as the mass peaks for both halos and the X-ray peak for the main cluster, produces highly consistent results ($< 0.2\sigma$). We report that the masses of the main and sub halos are $M_{200} = 5.85^{+1.30}_{-1.18} \times 10^{14}\ M_\odot$ ($2.22^{+0.80}_{-0.58} \times 10^{14}\ M_\odot$) and $M_{200} = 1.70^{+0.73}_{-0.59} \times 10^{14}\ M_\odot$ ($0.80^{+0.36}_{-0.25} \times 10^{14}\ M_\odot$), respectively, and the concentrations of the main and sub halos are $c_{200} = 3.84^{+0.04}_{-0.02}$ ($14.56^{+8.75}_{-4.72}$) and $c_{200} = 4.16^{+0.24}_{-0.11}$ ($14.62^{+5.11}_{-4.28}$), respectively, with (without) the $c$-$M$ relation. While there is approximately a factor of two systematic difference in the estimated masses, the mass ratio is approximately 3:1 for both cases. We further discuss the validity of the $c$-$M$ relation in the next section.

**Non-parametric Mass Estimation**

Because a somewhat large difference is found between the Perseus masses with and without the $c$-$M$ relation, it is important to compare the results with those obtained from non-parametric approaches. Here we choose to estimate the non-parametric mass of the Perseus cluster using the aperture mass densitometry (AMD) with the following integral[28,29]:

$$\zeta_c(r_1, r_2, r_{max}) = \bar{\kappa}(r \leq r_1) - \bar{\kappa}(r_2 < r \leq r_{max}) = 2\int_{r_1}^{r_2} \frac{\langle \gamma_T \rangle}{r} dr + \frac{2}{1-r_2^2/r_{max}^2} \int_{r_2}^{r_{max}} \frac{\langle \gamma_T \rangle}{r} dr, \quad (1)$$

where $\langle \gamma_T \rangle$ is the azimuthally averaged tangential shear, $r_1$ is the aperture radius, and $r_2$ and $r_{max}$ are the inner- and the outer-radii of the annulus. We estimate the density contrast between the inner ($r < r_1$) and outer ($r_2 < r < r_{max}$) regions. For the control annulus, we choose $r_2 = 92'$ (2 Mpc) and $r_{max} = 95'$ (2.1 Mpc) which is derived solely from the SDSS data. The predicted density within the annulus is $\bar{\kappa} = 0.0017$ from the single NFW halo fitting with the $c$-$M$ relation. We update the convergence ($\kappa$) iteratively to determine the aperture mass as our input shear is the reduced shear.

Figure 3 shows the AMD result compared with the best-fit single NFW models based on different assumptions regarding the $c$-$M$ relation. The projected mass profile of the AMD is consistent with the NFW aperture mass obtained with the $c$-$M$ relation, suggesting that the assumption on the $c$-$M$ relation better fits to the Perseus' halo profile. The AMD projected mass at the 1 Mpc (2 Mpc) radius is $6.4 \pm 2.5 \times 10^{14}\ M_\odot$ ($8.8 \pm 6.3 \times 10^{14}\ M_\odot$). The large uncertainty at the 2 Mpc radius arises because the measurement there primarily relies on shear data from the SDSS (see Figure 2 and Equation 1), which provides a source density 30 times lower than that of the HSC data. In contrast, parametric estimation, based on $\chi^2$ minimization, is less sensitive to additional noisy data and does not lead to significantly larger errors when such data are incorporated.

**Mass bridge**

The mass bridge was detected with the integrated significance of 3.2σ (Extended Data Figure 2). This feature is also supported by the recent Euclid study reporting significant elongation of the distributions in intracluster light and intracluster globular clusters from the Perseus core toward the west[30]. We estimate masses of the bridge by simultaneously fitting a cylindrical bridge model having a length of 430 kpc and a characteristic width of 0.1 Mpc between the two NFW halo profiles using the MCMC analysis. We fix the center of the profiles at each halo's brightest galaxy and assume the *c-M* relation. We obtain a bridge's mass density of $\rho = 123^{+97}_{-76}\rho_{bkg}$ in units of the background density at the Perseus redshift. With the inclusion of the mass bridge in our fitting, the main (sub) halo mass is estimated to be $M_{200} = 6.65^{+2.04}_{-1.74} \times 10^{14} M_\odot$ ( $2.12^{+1.23}_{-0.91} \times 10^{14} M_\odot$ ). The linear mass density of the mass bridge is estimated to be $0.24^{+0.18}_{-0.15} \times 10^{14} M_\odot$ Mpc$^{-1}$.

**Interpretation**

A mass bridge can arise from dynamical friction when two clusters pass through each other, as is often found in numerical simulations and observations[30]. With the current WL masses and the timing argument[31], we estimate the initial infall velocity to be ~1,300 km s$^{-1}$ when the separation between the two halos is 2 Mpc. If the merger is head-on in the plane of the sky, the time since collision at the current observed epoch is less than 0.3 Gyrs. This recent merger scenario seems unlikely because the severely disturbed core cannot be restored to the current near-equilibrium state observed in the central region (<60 kpc)[32] within such a short time scale. To reconcile the features at the Perseus core and the large cold front observed ~700 kpc east, Bellomi et al.[33] suggest a 1:5 major merger with a large impact parameter. According to this scenario, the large eastern cold front is 6 to 8.5 Gyrs old, and the subcluster has completely merged into the main cluster.

We carried out idealized magneto-hydrodynamic (MHD) simulations of the Perseus cluster in a similar fashion, but with our WL masses (see Methods for details). We find that a ~3:1 off-axis merger can reproduce the large eastern cold front and the mass bridge at the third apocenter, which is ~5.5 Gyr after the first core crossing (Figure 4). Unlike the result of Bellomi et al.[33], the subcluster is still intact ~400 kpc west of the main cluster core while the density of the mass bridge increases with time (Figure 4). The expected linear mass density of the bridge is ~$0.5 \times 10^{14}$ $M_\odot$ Mpc$^{-1}$ at the third apocenter, which is roughly consistent with the current measurement (~$0.2 \times 10^{14}$ $M_\odot$ Mpc$^{-1}$).

Although the cold fronts in the Perseus cluster have long been believed to originate from cluster collisions, no merging companion has been identified to date. This limitation leads us to a scenario wherein the responsible subcluster has already merged with the main cluster[33]. The discovery of the subcluster and mass bridge by WL not only provides direct evidence that the Perseus cluster has undergone a major merger but also allows us to constrain the merging scenario in detail. Furthermore, the alignment of the substructures effectively rules out the possibility that they might be unrelated line-of-sight structures outside the Perseus cluster (Extended Data Figure 1).

One of the critical questions on the cold front study is how these features have maintained their characteristic density structure over cosmic timescales. Leading theories[34,35] suggest that magnetic fields play an essential role in preserving the structures. Consequently, detailed studies of cold fronts have the potential to infer the strength of the cosmic magnetic field through careful numerical simulations[8,33]. The improved understanding of the merging history of the Perseus cluster enabled by the current study contributes to such efforts by providing context for interpreting the observational data and refining theoretical models.

## Methods

**Data Reduction**

The Perseus cluster was observed with the Subaru/HSC with *g*, *r*, and *i* filters in 2014. Supplementary Table 1 summarizes the archival data that we retrieved from the SMOKA portal (https://smoka.nao.ac.jp/). We performed the low-level CCD processing (i.e., dark, overscan subtraction, flat fielding and etc.) with the the Rubin Telescope Legacy Survey of Space Time (LSST) software stack[36] v22_0_0 and updated the astrometric header information to the TPV format using the sip_to_pv (https://github.com/stargaser/sip_tpv.git) script[37]. We created mosaic science and weight images for each filter using the calibrated frames with the SWARP package[38] (http://www.astromatic.net/software/swarp). For the object detection and photometry, we ran Sextractor[39] (https://github.com/astromatic/sextractor) with the dual image mode on the *gri* combined detection image across the given filters. The photometric calibration is performed relative to the SDSS DR16 catalog[40] with the cross-matched objects in the same field. Readers are referred to Finner et al.[41] and HyeongHan et al.[42] for more details.

We encapsulated time- and spatial-dependent point-spread function (PSF) by performing a principal component analysis (PCA) of the observed stars for each resampled frame[43,44]. We reconstructed 2D PSF models by stacking the interpolated PSF image at the location of source galaxies for all associated resampled frames[16,27,42]. As a measure of remaining systematics arising from the imperfect PSF modeling, we calculated the auto- and cross-correlation functions of the PSF residual (observed star ellipticity - model PSF ellipticity). The amplitudes are below $10^{-6}$ on angular scale greater than $1'$ for *gri* images, indicating a negligible effect on our analysis.

The Sloan Digital Sky Survey[15] (SDSS) is a large (8000 sq. degree), multi-band[45] (*ugriz*) imaging and spectroscopic survey conducted by a 2.5-m telescope at Apache Point Observatory.

The Perseus cluster was observed in 2003 and 2005 in a drift-scan mode[46]. We retrieved astrometrically[47] and photometrically[48,49] calibrated science frames within 3 × 3 sq. degree field of view centered at NGC 1275 from the Science Archive Server (SAS; https://data.sdss.org/sas/). We generated mosaic images (Supplementary Figure 3) and detected sources for *gri* filters as of the Subaru/HSC imaging data.

We retrieved PSF metadata files from the SAS that were constructed by the PCA analysis to capture the PSF variation in temporal and spatial domain[50,51]. We generated PSF models for each calibrated frame at a desired location so that we obtained the final PSF models as in the HSC data analysis. The residual correlation functions show the amplitude below $10^{-5}$ on angular scale greater than $1'$ for *gri* images.

**Shear Measurement**

For both HSC and SDSS data, we measure the shape of source galaxies with a forward modeling approach[16,27,42]. In this study, we fit the PSF-convolved elliptical Gaussian model $[M_{j,k}(x,y)]$ to the observed postage image $[O_{j,k}(x,y)]$ across three different filters (i.e., *gri*) simultaneously[52]:

$$\chi^2 = \sum_{j=1}^{n} \sum_{k}^{g,r,i} \frac{[O_{j,k}(x,y) - M_{j,k}(x,y)]^2}{\sigma_{j,k}^2}, \qquad (2)$$

where $\sigma_{j,k}$ is a root mean square (rms) postage image. For neighboring objects, we assign $10^{10}$ to the rms value based on the segmentation map obtained from the SExtractor. The minimization is performed by using the MPFIT module[53] with three shape parameters (i.e., ellipticity, semi-minor axis, and position angle), three normalization parameters (one per filter) while fixing galaxy position, noise, and background level.

We select background source galaxies using photometric and shape criteria. For the HSC data, the objects having magnitudes between $23 < r < 26$ are chosen without a color cut since the member galaxy and foreground contamination is negligible thanks to the proximity ($z = 0.0179$) of the Perseus cluster[16]. Based on a comparison with external photometric redshift catalogs, we estimate that the Perseus cluster member contamination is less than 1% within the selection window. We apply size criteria to ensure that the half-light radius of the source in the $i$ band is greater than $0''.4$, which corresponds to the half-light radius of the stellar objects. In addition, the semi-minor axis is required to be greater than 0.4 pixels to remove artifacts caused by pixelation effect. Without this additional measure, the ellipticity distribution shows a non-isotropic, characteristic pattern (see Figure 14 in Jee et al.[54]). By imposing the STATUS = 1 criterion[53], we discard objects if the MPFIT fitting status is reported to be less than ideal. The ellipticity measurement error is required to be less than 0.3 to avoid objects with measurement errors exceeding the intrinsic shape dispersion. The resulting source density is ~22.7 per sq. arcmin, corresponding to ~47,500 per sq. Mpc at the redshift of the Perseus cluster. This high source density per physical area enables the detection of low-contrast substructures. For comparison, the same angular source density (~22.7 per sq. arcmin) would correspond to only ~169 per sq. Mpc (i.e., reduction by a factor of ~280) if the lens were at $z = 0.5$. After accounting for the difference in the lensing efficiency and the correlated noise from large-scale structures, the net S/N increase is approximately a factor of 3.

For the SDSS data, we select sources with magnitudes between $18 < r < 23$ in the region outside the HSC coverage. Although the magnitude interval is substantially brighter than in the case of the HSC data, we were able to avoid the cluster member contamination utilizing the spectroscopic and photometric redshift information[55]. To minimize the stellar contamination, we select objects having *i* band half-light radius greater than the value of stellar objects. Similar to the HSC case, we impose the ellipticity error cut <0.4, the semi-minor axis cut >0.6 pixel, and the STATUS=1 requirement. The final source density for the SDSS imaging data is 0.6 arcmin$^{-2}$ or 1,350 Mpc$^{-2}$. Since we rely on the SDSS data only for the region outside the HSC coverage, the combination of the two catalogs is straightforward and does not include duplicated objects.

In the absence of redshift information for each source galaxy, we estimate the source redshift using a photometric redshift as a reference catalog. We apply the magnitude constraint to the COSMOS2020 catalog[56] and weight the number density in each magnitude bin to match the source catalog to account for the difference in the imaging depth. The lensing efficiency is $\langle\beta\rangle = 0.95$ (corresponding effective redshift is $z_{eff} = 0.43$) and the width of the distribution is $\langle\beta^2\rangle = 0.92$. We correct the reduced shear, $g = \gamma/(1-\kappa)$ where $\gamma$ is the shear and $\kappa$ is the convergence, in a first order to account for the redshift distribution of the sources[57]: $g' = [1 + \kappa(\langle\beta^2\rangle/\langle\beta\rangle^2 - 1)]g$.

The Perseus cluster ($b \sim -13°$) is located near the Galactic plane where the dilution effect by stellar contamination can be as high as 5%[58,59]. Accounting for this effect, we apply the multiplicative bias correction $m = 1.1$[58] to the HSC shear catalog.

**Idealized Simulation Setup**

We run idealized MHD cluster merger simulations using a GPU-accelerated Adaptive Mesh Refinement code, GAMER-2[60], to investigate the evolution of dark matter distribution between halos during a cluster merger. The ideal MHD equations are solved using the corner transport upwind methods[61] (CTU) and HLLD Riemann solver[62], while additional physics such as cooling, and feedback are neglected for simplicity. We implement the magnetic field since it suppresses the development of the Kelvin-Helmholtz instability, which contributes to the formation of a cold front at a late merger phase[63].

The initial conditions for the clusters were generated using the cluster_generator package (https://github.com/jzuhone/cluster_generator). The code randomly distributes dark matter particles to follow the NFW profile and computes their velocity at assigned positions using the Eddington Formula. The intracluster medium (ICM) density is assigned to each cell with a modified beta profile[64], as described by the following equation:

$$n_p n_e = n_0 \frac{(r/r_c)^{-\alpha}}{(1+r^2/r_c^2)^{3\beta-\alpha/2}} \frac{1}{\left(1+r^\gamma/r_{s,g}^\gamma\right)^{\epsilon/\gamma}}, \quad (3)$$

where $n_p$ and $n_e$ are the number densities of the protons and electrons, respectively. We use the core radius $r_c = 0.06 r_{2500}$, the scale radius $r_{s,g} = 0.3 r_{200}$, and the slope $(\alpha, \beta, \gamma, \epsilon) = (1.0, 0.67, 3.0, 3.0)$ to design cool core cluster. The profiles are defined up to 5 Mpc from the cluster center. A turbulent magnetic field is initialized in each cell by generating a Gaussian random field in Fourier space, with amplitudes following the Kolmogorov power spectrum. After transforming a random magnetic field into the physical space, the magnetic field magnitude is rescaled to have an average ratio of thermal and magnetic pressure of ~100.

Based on our WL mass estimate, the total mass of the main and sub-cluster is set as $5.8 \times 10^{14}\ M_\odot$ and $1.7 \times 10^{14}\ M_\odot$, respectively with concentration parameters of $c = 3.8$ and $4.2$, following the *c-M* relation. The clusters are placed in a simulation box of 15 Mpc-size with an initial separation of 3 Mpc. We set a fixed number of cells ($8^3$) for each "patch", which consists of fixed number of cells that are adaptively refined when the number of dark matter particles exceeds a threshold. It allows us to achieve the spatial resolution of the grid down to ~8 kpc near the cluster centers and ~200 kpc at the outskirts, based on the number of dark matter particles. We perform merger simulations with varying initial velocities (1,000-1,800 km/s) and impact parameters (0.5-

1.8 Mpc). Here, we present the result of an off-axis cluster merger with an initial velocity of 1200 km/s and a large impact parameter (1.8 Mpc), following the merger scenario proposed in previous studies (e.g., Bellomi et al. 2023). Readers are referred to Brzycki & ZuHone[65] and Lee et al. [66] for more details.

**Data Availability:** The raw Subaru/HSC and SDSS imaging data used for the current study are publicly available. The processed mosaic images and data points in the article figures are available on request from the authors.

**Code Availability:** Our custom data processing codes are available on request from the authors.

**Acknowledgements:**

M.J.J. acknowledge support for the current research from the National Research Foundation (NRF) of Korea under the programs 2022R1A2C1003130 and RS-2023-00219959. WL acknowledges support from the National Research Foundation of Korea(NRF) grant funded by the Korea government(MSIT) (RS-2024-00340949). H.S.H. acknowledges the support of the National Research Foundation of Korea (NRF) grant funded by the Korea government (MSIT), NRF-2021R1A2C1094577, Samsung Electronic Co., Ltd. (Project Number IO220811-01945-01), and Hyunsong Educational & Cultural Foundation.

**Author contributions:**

K.H. designed the research, reduced the data, measured the WL signals, and wrote the manuscript. M.J.J. designed the research, developed the reduction pipeline, and wrote the manuscript. W.L.

performed the numerical simulations, reconstructed the merger scenario, and wrote the manuscript. J.Z. contributed to the numerical simulations and their interpretation. I.Z. contributed to the interpretation of the observations. W.K. and H.S.H provided the spectroscopic data and contributed to the discussion.

**Competing interests:**

The authors declare no competing interests.

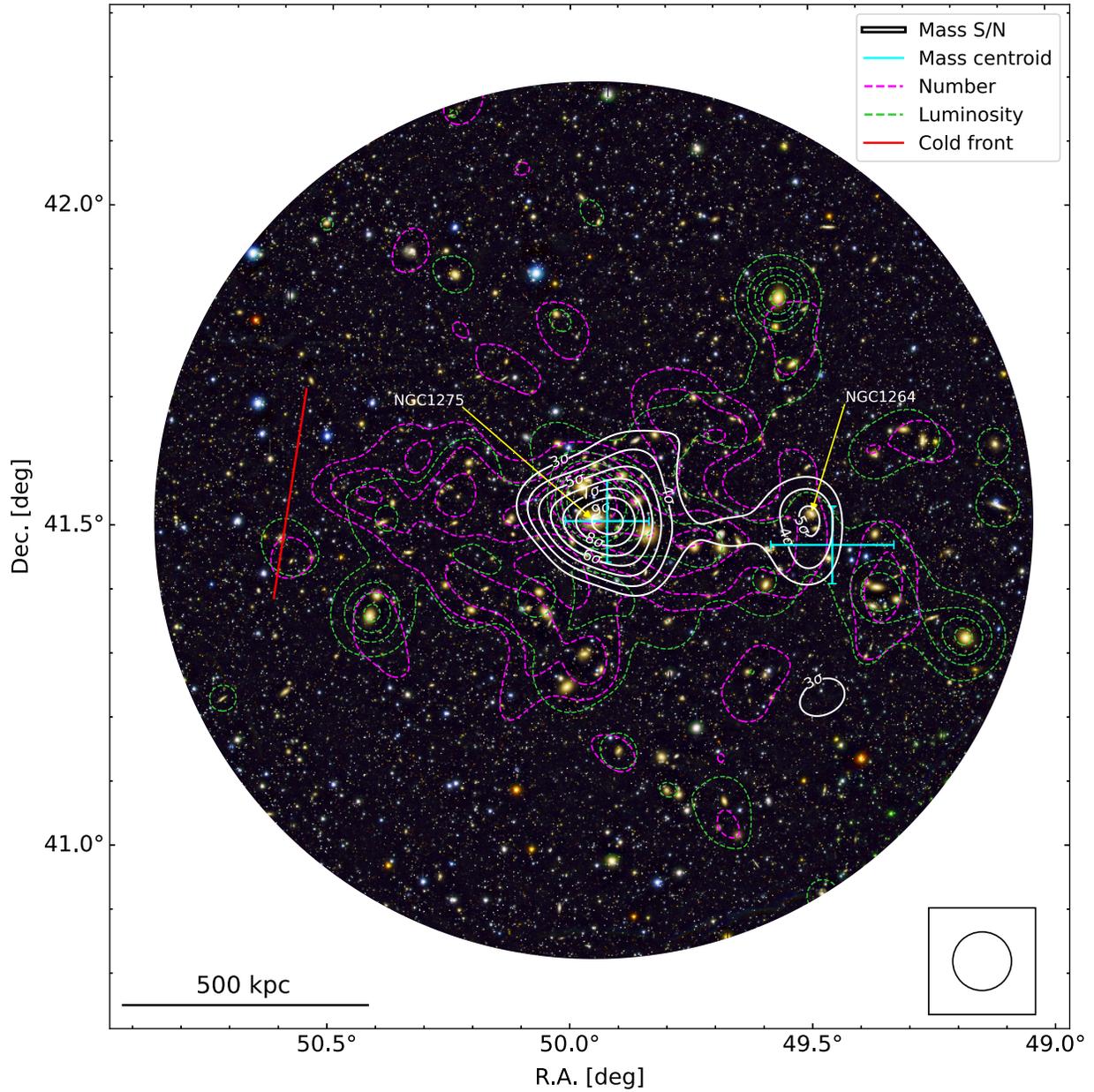

**Figure 1: Dark matter and galaxy distribution of the Perseus cluster.** The RGB color image consists of blue, green, and red channels represented by intensities in the Subaru/HSC *g*, *r*, and *i*-band images, respectively. White contours indicate a significance ($\kappa/\sigma_\kappa$) of the reconstructed mass that starts from $3\sigma$ with $1\sigma$ step each. We reconstructed a convergence ($\kappa$) map by using the Fourier-inversion method[67] and obtained its noise ($\sigma_\kappa$) map by generating 1000 convergence fields from the resampled HSC shear catalogs. Cyan indicates the $1\sigma$ uncertainties of the mass centroid.

The circle in the lower right corner represents the effective mass reconstruction smoothing scale $\sigma = 165''$. We obtained consistent results with other mass reconstruction algorithms[68,69]. Magenta (green) contours show the smoothed number (luminosity) distribution of spectroscopically confirmed member galaxies identified in Kang et al.[17]. The red solid line indicates the location of the large cold front ~700 kpc east[7,8]. Throughout the paper, we assume a flat $\Lambda$CDM cosmology characterized by $h = 0.7$ and $\Omega_{m,0} = 1 - \Omega_{\Lambda,0} = 0.3$.

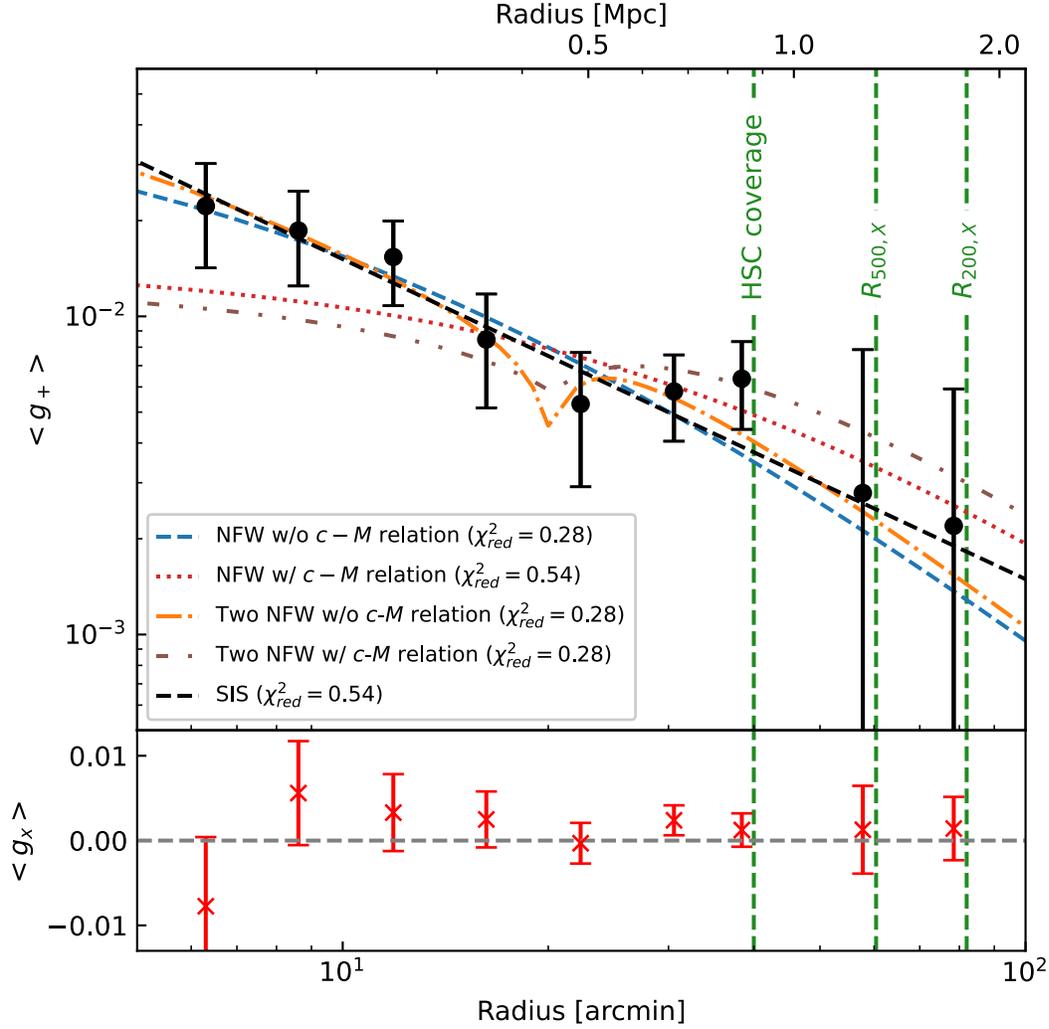

**Figure 2: Reduced shear profile of the Perseus cluster using the HSC-SDSS merged shear catalog.** Top: Azimuthally averaged tangential shear profile with respect to the BCG. Each error bar indicates $1\sigma$ confidence level that accounts for the ellipticity dispersion and measurement error. Green dashed lines represent the $R_{500}$ and $R_{200}$ adopted from Simionescue et al.[21]. Dashed lines are the best-fit models with different assumptions. Readers are referred to the main text for more details. Bottom: Cross shear signal obtained by rotating galaxy position angles by 45°. The cross shear is consistent with the null signal.

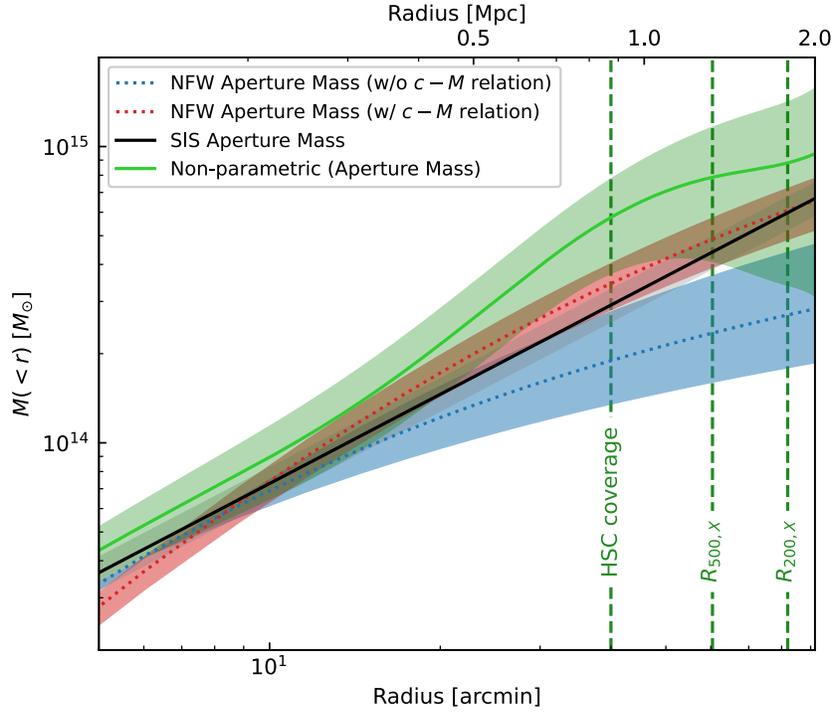

**Figure 3: Projected mass profiles comparing parametric (analytic projection of NFW profiles) results with nonparametric (aperture mass densitometry) ones.** The red (blue) dotted line shows an analytic projection of the NFW profile with (without) the *c-M* relation. The green solid line represents the aperture mass densitometry. The black solid line indicates the aperture mass of the SIS profile. Shaded regions indicate $1\sigma$ statistical uncertainties. Dashed vertical green lines indicate the radius where the HSC observation encloses and the radii $R_{500}$ and $R_{200}$ estimated by the X-ray observation[21] where the average density becomes 500 and 200 times the critical density of the universe, respectively. The mass difference at $R_{200}$ between the NFW fitting results obtained with and without the c-M relation is approximately a factor of two. The non-parametric mass profile favors the NFW result obtained with the c-M relation within the HSC field.

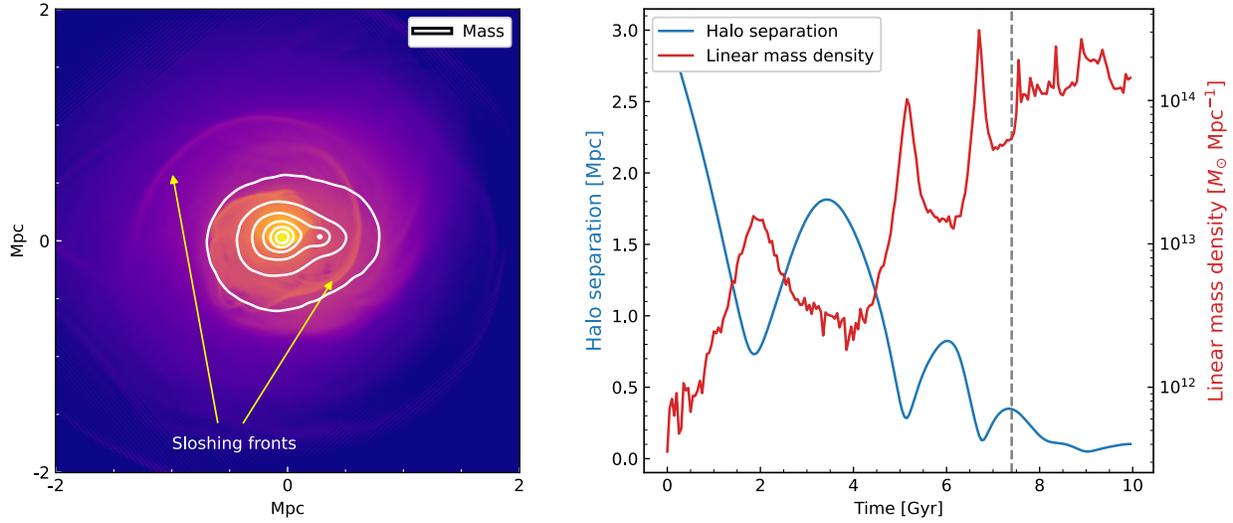

**Figure 4: Numerical simulation of the hypothesized merger in the Perseus cluster.** Left: Snapshot of the dark matter density contours (white) overlaid on the Gaussian-gradient magnitude (σ =0.125 kpc) filtered X-ray surface brightness map at ~5.5 Gyr after the first core passage. The dark matter map is smoothed with a σ =10 kpc Gaussian kernel. The X-ray surface brightness map is produced from the 0.5-7 keV energy interval. Right: Halo separation (blue) and mass bridge density (red) evolution with time. The mass bridge linear density is measured from a 30 kpc length and a 100 kpc width cylindrical volume at the center of the two halos. The vertical dashed line indicates the 3$^{rd}$ apocenter phase displayed in the left panel. The initial mass of the main (sub) clusters is $5.8 \times 10^{14}\ M_\odot$ ($1.7 \times 10^{14}\ M_\odot$), adopting the value from the WL mass obtained with the *c-M* relation. While we did not finetune our numerical simulation to exactly replicate the observations, the current snapshot reasonably reproduces the observed X-ray and WL features. This includes the cold front ~700 kpc east of the main core, the subcluster ~430 kpc west of the main core, and the mass bridge between the main and sub clusters.

# References

1. Edge, A. C. et al. An X-ray flux-limited sample of clusters of galaxies: evidence for evolution of the luminosity function. *Mon. Not. R. Astron. Soc.* **245**, 559 (1990).
2. Fabian, A. C. & Nulsen, P. E. J. Subsonic accretion of cooling gas in clusters of galaxies. *Mon. Not. R. Astron. Soc.* **180**, 479-484 (1977)
3. Soboleva, N. S. et al. 3C84 – the 5' Radio Halo and a Search for Scattering of the Emission of the Compact Central Source by the Intracluster Gas. *Sov. Astron. Lett.*, **9**, 305 (1983).
4. Gendron-Marsolais, M. et al. Deep 230-470 MHz VLA observations of the mini-halo in the Perseus cluster. *Mon. Not. R. Astron. Soc.* **469**, 3872-3880 (2017).
5. Churazov, E., et al. XMM-Newton Observations of the Perseus Cluster. I. The Temperature and Surface Brightness Structure. *Astrophys. J.* **590**, 225 (2003).
6. Fabian, A. C. et al. A wide Chandra view of the core of the Perseus cluster. *Mon. Not. R. Astron. Soc.* **418**, 2154-2164 (2011).
7. Simionescu, A. et al. Large-scale Motions in the Perseus Galaxy Cluster. *Astrophys. J.* **757**, 182 (2012).
8. Walker, S. A. et al. The split in the ancient cold front in the Perseus cluster. *Nature Astronomy.* **2**, 292-296 (2018).
9. Brunzendorf, J. & Meusinger, H. The galaxy cluster Abell 426 (Perseus). A catalog of 660 galaxy position, isophotal magnitudes and morphological types. *Astron. Astrophys. Suppl.* **139**, 141-161 (1999).
10. Aguerri, J. A. L. et al. Deep spectroscopy in nearby galaxy clusters – V. The Perseus cluster. *Mon. Not. R. Astron. Soc.* **494**, 1681-1692 (2020).
11. Furusawa, H. et al. The on-site quality-assurance system for Hyper Suprime-Cam: OSQAH. *Publ. Astron. Soc. Jap* **70**, S3 (2018).
12. Kawanomoto, S. et al. Hyper Suprime-Cam: Filters. *Publ. Astron. Soc. Jap* **70**, 66 (2018).
13. Komiyama, Y. et al. Hyper Suprime-Cam: Camera dewar design. *Publ. Astron. Soc. Jap* **70**, S2 (2018).
14. Miyazaki, S. et al. Hyper Suprime-Cam: System design and verification of image quality. *Publ. Astron. Soc. Jap* **70**, S1 (2018).
15. York, D. G. et al. The Sloan Digital Sky Survey: Technical Summary. *Astrophys. J.* **120**, 1579 (2000).
16. HyeongHan, K. et al. Weak-lensing detection of intracluster filaments in the Coma cluster. *Nature Astronomy*. **8**, 377-383 (2024).
17. Kang, W. et al. A Deep Redshift Survey of the Perseus Cluster: Spatial Distribution and Kinematics of Galaxies. *Astrophys. J. Sppl.* **272**, 22 (2024).
18. Navarro, J. F. et al. The Structure of Cold Dark Matter Halos. *Astrophys. J.* **462**, 563 (1996).
19. Navarro, J. F. et al. A Universal Density Profile from Hierarchical Clustering. *Astrophys. J.* **490**, 493 (1997).
20. Ishiyama, T. et al. The Uchuu simulations: Data Release 1 and dark matter halo concentrations. *Mon. Not. R. Astron. Soc.* **506**, 4210-4231 (2021).
21. Simionescu, A. et al. Baryons at the Edge of the X-ray-Brightest Galaxy Cluster. *Sci*. **331**, 1576 (2011).
22. Girardi, M. et al. Optical Mass Estimates of Galaxy Clusters. *Astrophys. J.* **505**, 74 (1998).
23. Khanday, S. A. et al. Morphology, colour-magnitude, and scaling relations of galaxies in Abell 426. *Mon. Not. R. Astron. Soc.* **515**, 5043-5061 (2022).

24. Jee, M. J. et al. Weighing "El Gordo" with a Precision Scale: Hubble Space Telescope Weak-lensing Analysis of the Merging Galaxy Cluster ACT-CL J0102-4915 at z=0.87. *Astrophys. J.* **785**, 20 (2014).
25. Finner, K. et al. MC$^2$: Subaru and Hubble Space Telescope Weak-lensing Analysis of the Double Radio Relic Galaxy Cluster PLCK G287.0+32.9. *Astrophy. J.* 851, 46 (2017).
26. Weinmann, S. M. et al. Dwarf galaxy populations in present-day galaxy clusters – I. Abundances and red fractions. *Mon. Not. R. Astron. Soc.* **416**, 1197-1214 (2011).
27. HyeongHan, K. et al. Discovery of a Radio Relic in the Massive Merging Cluster SPT-CL J2023-5535 from the ASKAP-EMU Pilot Survey. *Astrophy. J.* 900, 127 (2020).
28. Fahlman, G. et al. Dark Matter in MS 1224 from Distortion of Background Galaxies. *Astrophys. J.* **437**, 56 (1994).
29. Clowe, D. et al. Weak Lensing by High-Redshift Clusters of Galaxies. I. Cluster Mass Reconstruction. *Astrophys. J.* **539**, 540 (2000).
30. Cha, S. et al. Precision MARS Mass Reconstruction of A2744: Synergizing the Largest Strong-lensing and Densest Weak-lensing Data Sets from JWST. *Astrophys. J.* **961**, 186 (2024).
31. Sarazin, C. L. The Physics of Cluster Mergers. *Astrophys. Space Science Library.* **272**, 1-38 (2002).
32. Hitomi Collaboration et al. The quiescent intracluster medium in the core of the Perseus cluster. *Nature*, **535**, 117-121 (2016).
33. Bellomi, E. et al. On the Origin of the Ancient, Large-Scale Cold Front in the Perseus Cluster of Galaxies. arXiv:2301.09422 (2023).
34. Lyutikov, M. Magnetic draping of merging cores and radio bubbles in clusters of galaxies. *Mon. Not. R. Astron. Soc.* **373**, 73-78 (2007).
35. Zuhone, J., et al. Sloshing of the Magnetized Cool Gas in the Cores of Galaxy Clusters. *Astrophys. J.* **743**, 16 (2011).
36. Bosch, J. et al. The Hyper Suprime-Cam software pipeline. *Publ. Astron. Soc. Jap* **70**, S5 (2018).
37. Shupe, D. L. et al. More flexibility in representing geometric distortion in astronomical images. *Software and Cyberinfrastructure for Astronomy II.* Proc. SPIE **8451**, 84511M (2012).
38. Gruen, D. et al. Implementation of Robust Image Artifact Removal in SWarp through Clipped Mean Stacking. *Publ. Astron. Soc. Pac.* **126**, 158 (2014).
39. Bertin, E., & Arnouts, S. SExtractor: Software for source extraction. *Astron. Astrophys*. **117**, 393-404 (1996).
40. Ahumada, R. et al. The 16[th] Data Release of the Sloan Digital Sky Surveys: First Release from the APOGEE-2 Southern Survey and Full Release of eBOSS Spectra. *Astrophys. J. Suppl.* **249**, 3
41. Finner, K. et al. Exemplary Merging Clusters: Weak-lensing and X-Ray Analysis of the Double Radio Relic, Merging Galaxy Clusters MACS J1752.0+4440 and ZWCL 1856.8+6616. *Astrophy. J.* 918, 72 (2021).
42. HyeongHan, K. et al. Weak-lensing Analysis of the Complex Cluster Merger A746 with Subaru/Hyper Suprime-Cam. *Astrophys. J.* **962**, 100 (2024).
43. Jee, M. J. et al. Principal Component Analysis of the Time- and Position-dependent Point-Spread Function of the Advanced Camera for Surveys. *Publ. Astron. Soc. Pac.* 199, 1403 (2007).
44. Jee, M. J. et al. Toward Precision LSST Weak-Lensing Measurement. I. Impacts of Atmospheric Turbulence and Optical Aberration. *Publ. Astron. Soc. Pac.* 123, 596 (2011).

45. Fukugita, M. et al. The Sloan Digital Sky Survey Photometric System. *Astrophy. J.* **111**, 1748 (1996).
46. Stoughton, C. et al. Sloan Digital Sky Survey: Early Data Release. *Astrophy. J.* **123**, 485 (2002).
47. Pier, J. R. et al. Astrometric Calibration of the Sloan Digital Sky Survey. *Astrophy. J.* **125**, 1559 (2003).
48. Hogg, D. W. et al. A Photometricity and Extinction Monitor at the Apache Point Observatory. *Astrophy. J.* **122**, 2129 (2001).
49. Tucker, D. L. et al. The Sloan Digital Sky Survey monitor telescope pipeline. *Astronomische Nachrichten*. **327**, 821 (2006).
50. Lupton, R. et al. The SDSS Imaging Pipelines. *Astron. Soc. Pacif. Conf. Ser.,* **238**, 269 (2001).
51. Sheldon, E. S. et al. The Galaxy-Mass Correlation Function Measured from Weak-Lensing in the Sloan Digital Sky Survey. *Astrophy. J.* **127**, 2544 (2004).
52. Ahn, E. et al. Substructures within Substructures in the Complex Post-Merging System A514 Unveiled by High-Resolution Magellan/Megacam Weak Lensing. arXiv:2404.04321 (2024).
53. Markwardt, C. B. Non-linear Least Squares Fitting in IDL with MPFIT. *Astron. Soc. Pacif. Conf. Ser.,* **411**, 251-254 (2009).
54. Jee, M. J. et al. Cosmic Shear Results from the Deep Lens Survey. I. Joint Constraints on $\Omega_M$ and $\sigma_8$ with a Two-dimensional Analysis. *Astrophys. J.* **765**, 74 (2013).
55. Beck, R. et al. Photometric redshifts for the SDSS Data Release 12. *Mon. Not. R. Astron. Soc.* **460**, 1371-1381 (2014)
56. Weaver, J. R. et al. COSMOS2020: A Panchromatic View of the Universe to z ~ 10 from Two Complementary Catalogs. *Astrophys. J. Suppl.* **258**, 11 (2022).
57. Seitz, C. & Schneider, P. Steps towards nonlinear cluster inversion through gravitational distortions. III. Including a redshift distribution of the sources. *Astron. Astrophys*. **318**, 687-699 (1997).
58. Jee, M. J. et al. MC$^2$: Constraining the Dark Matter Distribution of the Violent Merging Galaxy Cluster CIZA J2242.8+5301 by Piercing through the Milky Way. *Astrophys. J.* **802,** 46 (2015).
59. Hoekstra, H. et al. A study of the sensitivity of shape measurements to the input parameters of weak-lensing image simulations. *Mon. Not. R. Astron. Soc*. **468,** 3295-3311 (2017).
60. Schive, H. et al. GAMER-2: a GPU-accelerated adaptive mesh refinement code – accuracy, performance, and scalability. *Mon. Not. R. Astron. Soc.* **481**, 4815-4840 (2018).
61. Colella, P. Multidimensional Upwind Methods for Hyperbolic Conservation Laws. *J. Computational Physics.* **87**, 171-200 (1990).
62. Miyoshi, T., & Kusano, K. A multi-state HLL approximate Riemann solver for ideal magnetohydrodynamics. J. Computational Physics. **208**, 315-344 (2005).
63. ZuHone, J. A., et al. Sloshing of the Magnetized Cool Gas in the Cores of Galaxy Clusters. *Astrophys. J.* **743**, 16 (2011).
64. Vikhlinin, A. et al. Chandra Sample of Nearby Relaxed Galaxy Clusters: Mass, Gas Fraction, and Mass-Temperature Relation. *Astrophys. J.* **640**, 691 (2006).
65. Brzycki, B. & ZuHone, J. A Parameter Space Exploration of Galaxy Cluster Mergers. II. Effects of Magnetic Fields. *Astrophys. J.* **883**, 22 (2019).
66. Lee, W. et al. Weak-lensing Mass Bias in Merging Galaxy Clusters. *Astrophys. J.* **945**, 71L (2023).
67. Kaiser, N., & Squires, G. Mapping the Dark Matter with Weak Gravitational Lensing. *Astrophys. J.* **404**, 441 (1993).

68. Fischer, P. & Tyson, J. A. The Mass Distribution of the Most Luminous X-ray Cluster RXJ 1347.5-1145 From Gravitational Lensing. *Astron. J.* 114, 14 (1997).
69. Cha, S. et al. MARS: A New Maximum-entropy-regularized Strong Lensing Mass Reconstruction Method. *Astrophys. J.* **931**, 127 (2022).

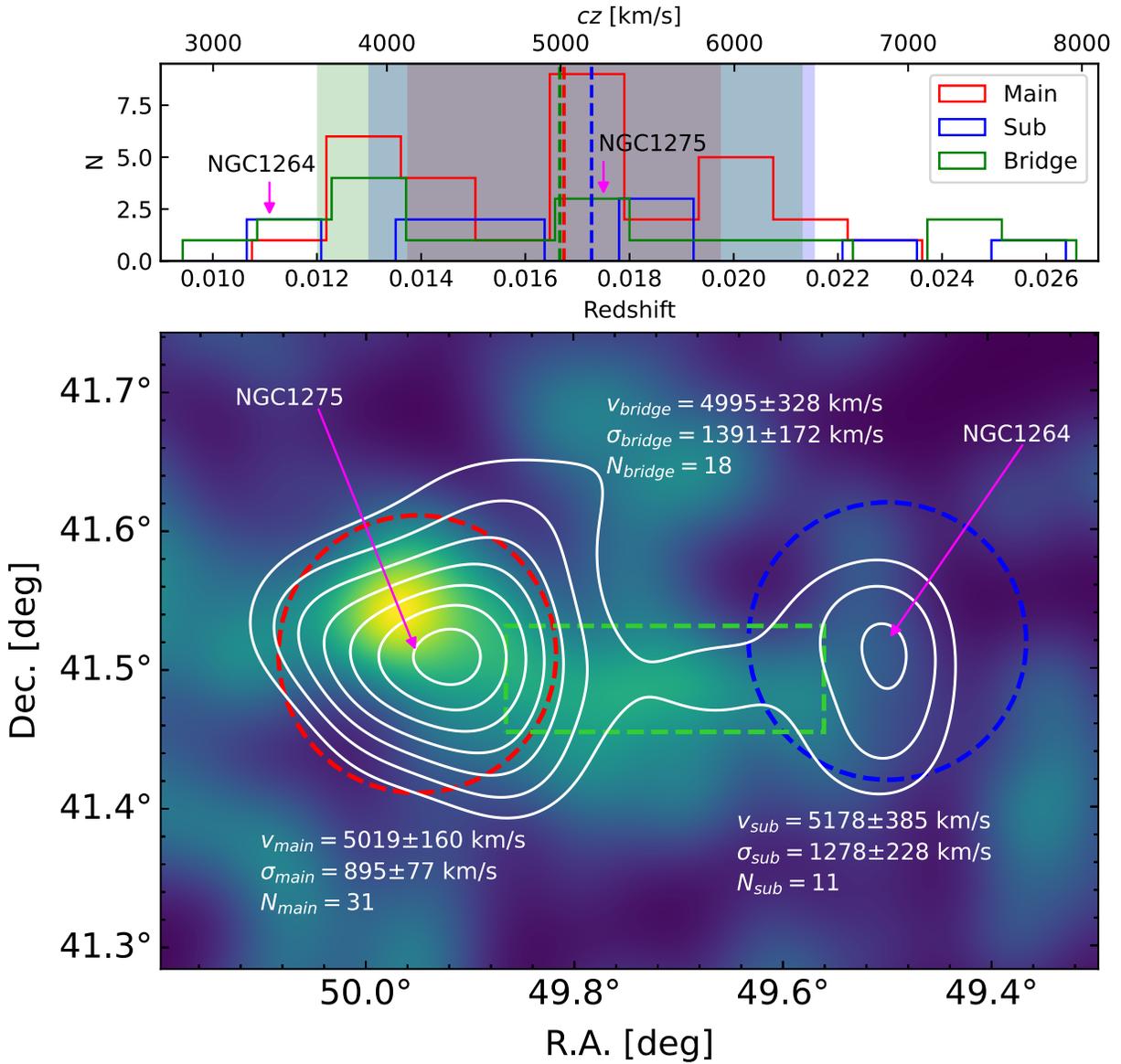

**Extended Data Figure 1:** Velocity structures of Perseus. Top: Velocity distributions of the cluster galaxies in the bridge, main, and sub halos are presented in green, red and blue, respectively. Vertical dashed lines indicate the average velocity of galaxies within each substructure while the shaded regions present the corresponding standard deviation. We impose slight velocity offsets (~30 km/s) to the histograms for the visibility. Magenta arrows mark the radial velocity of the associated brightest galaxies. Bottom: Background is a Perseus galaxy number density map

whereas white contours are the reconstructed mass significance. The outermost contour corresponds to 3$\sigma$ and the interval is 2$\sigma$. Red (main) and blue (sub) circles are centered at NGC 1275 and NGC 1264, respectively, with a radius of 6 arcmin (~130 kpc). Green box (300 kpc × 100 kpc) represents the bridge region. The mean LOS velocities of the cluster galaxies in the main halo, sub-halo, and bridge regions are estimated to be 5019±160 km/s, 5178±385 km/s, and 4995±328 km/s, respectively. This alignment rules out the possibility that the substructures might be unrelated LOS structures outside the Perseus cluster.

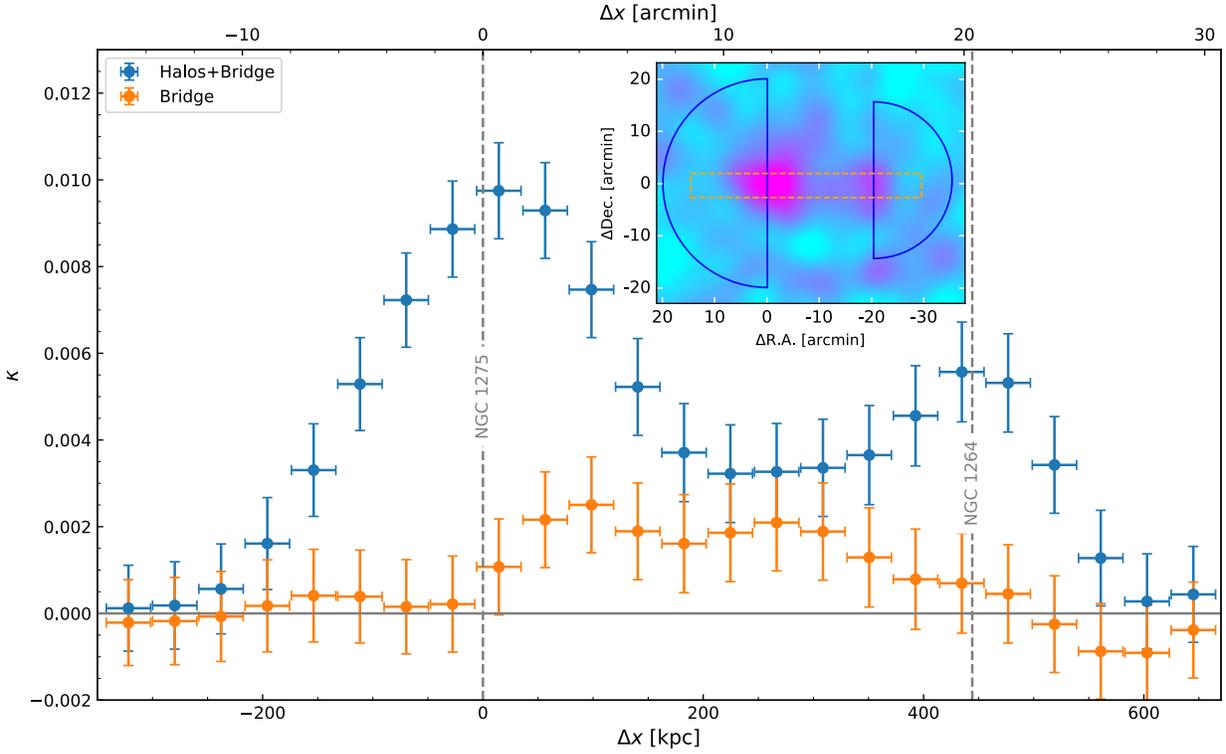

**Extended Figure 2:** Line profile along the mass bridge. The two blue half-circles (centered at NGC 1275 and NGC 1264) in the inset image indicates the areas from which we measure the radial density profiles of the two halos, while the orange rectangle represents the region used to estimate the line profile. In the main plot, the blue (orange) data points show the line profile without (with) the subtraction of the two halo profiles. The significance at the mid-point of the bridge drops from ~3σ to ~2σ when the two halo profiles are subtracted. Nevertheless, the integrated significance of the bridge within the segment bracketed by NGC 1275 and NGC 1264 remains 3.2σ. The error bars along the abscissa indicate a bin width and the 1σ statistical uncertainties along the ordinate.

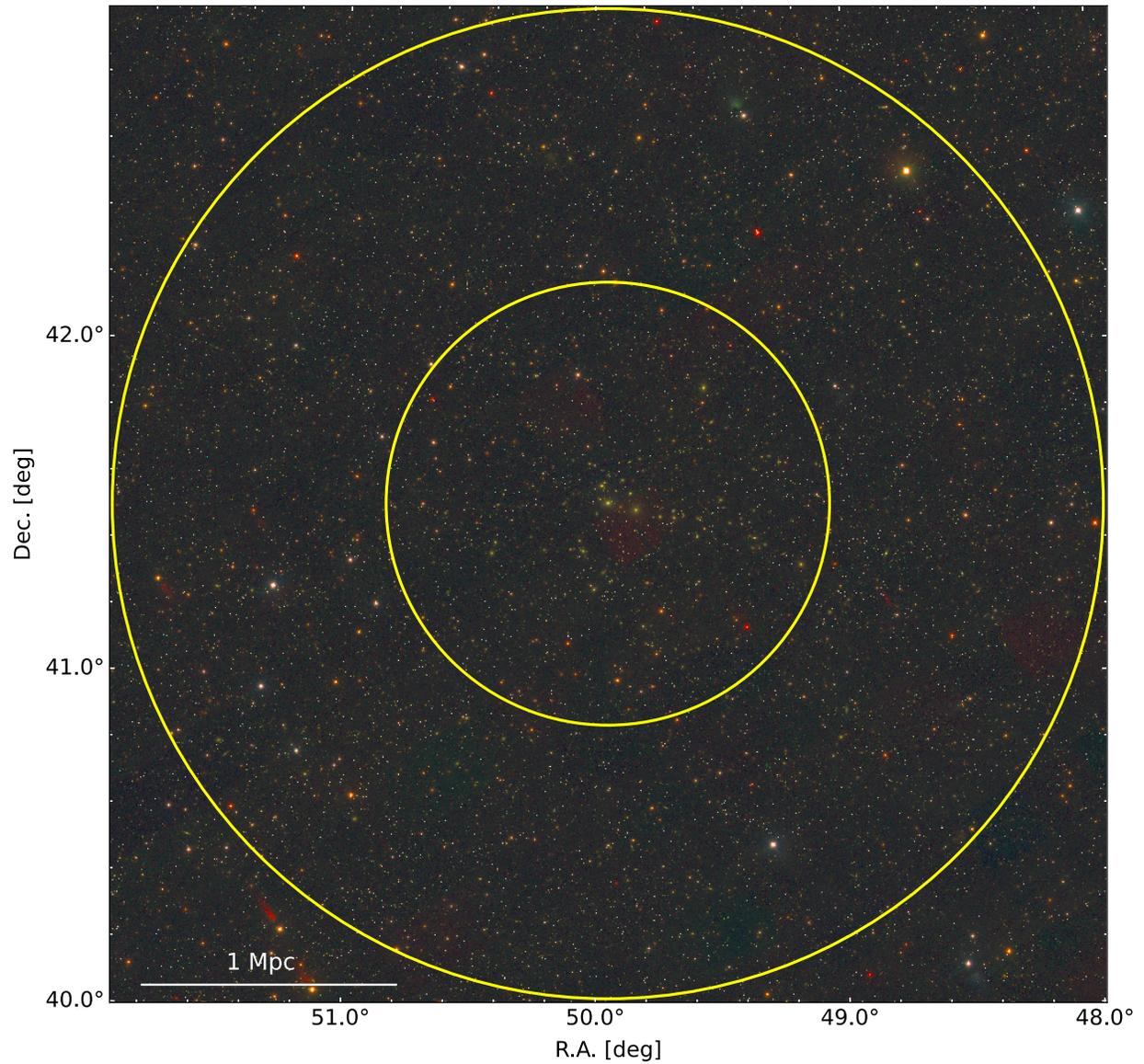

**Supplementary Figure 1: The color composite SDSS image.** The RGB color image consists of blue, green, and red channels represented by SDSS *g*, *r*, and *i*-band images, respectively. The outer yellow circle indicates the ~1.5° (2.1 Mpc) radius and the inner circle ($r = 40'$; ~870 kpc) represents the HSC coverage.

**Extended Data Table 1: Mass Estimates from best-fit models**

| Fitting Model | Main halo $M_{200}$ ($\times 10^{14} M_\odot$) | $c_{200}$ | Subhalo $M_{200}$ ($\times 10^{14} M_\odot$) | $c_{200}$ | $\chi^2/d.o.f.$ |
|---|---|---|---|---|---|
| Single-halo-fitting NFW w/ $c$-$M$ relation (1) | $6.82^{+1.76}_{-1.76}$ | $3.83^{+0.03}_{-0.01}$ | N/A | N/A | 0.54 |
| Single-halo-fitting w/o $c$-$M$ relation | $2.89^{+1.95}_{-0.98}$ | $10.61^{+7.84}_{-4.62}$ | N/A | N/A | 0.28 |
| Single-halo-fitting SIS (2) | $2.84^{+0.59}_{-0.52}$ | N/A | N/A | N/A | 0.54 |
| Two-halo-fitting NFW w/ $c$-$M$ relation (1) | $5.85^{+1.30}_{-1.18}$ | $3.84^{+0.04}_{-0.02}$ | $1.70^{+0.73}_{-0.59}$ | $4.16^{+0.24}_{-0.11}$ | 0.28 |
| Two-halo-fitting NFW w/o $c$-$M$ relation | $2.22^{+0.80}_{-0.58}$ | $14.56^{+8.75}_{-4.72}$ | $0.80^{+0.36}_{-0.25}$ | $14.62^{+5.11}_{-4.28}$ | 0.28 |

(1) We assume the $c$-$M$ relation adopted from Ishiyama et al.[20]. (2) We derive the mass from the best-fit velocity dispersion, $\sigma_{WL} = 808 \pm 53\ km\ s^{-1}$.

**Supplementary Table 1: Summary of HSC observations**

| Filter | Exposure Time (s) | Observation Date | ⟨FWHM⟩ (arcsec) | Limiting Magnitude (1) |
|---|---|---|---|---|
| HSC-$g$ | 23,850 | 2014 | 0.7 | 28.2 |
| HSC-$r$ | 4,380 | 2014 | 0.6 | 26.5 |
| HSC-$i$ | 2,190 | 2014 | 0.5 | 26.0 |

(1) We quote the $5\sigma$ limiting magnitude of a point source.